\documentclass[showpacs,superscriptaddress,twocolumn,aps]{revtex4} 
\usepackage{amsmath,graphicx} 
 
\begin{document} 
 
\title {Quantum Noise Measurement of a Carbon Nanotube Quantum Dot in the Kondo Regime} 
\author{J. Basset} 
\affiliation{Laboratoire de Physique des Solides, Univ. Paris-Sud, CNRS, UMR 8502, F-91405 Orsay Cedex, France.}  
\author{A.Yu. Kasumov} 
\affiliation{Laboratoire de Physique des Solides, Univ. Paris-Sud, CNRS, UMR 8502, F-91405 Orsay Cedex, France.}  
\author{C.P. Moca}
\affiliation{BME-MTA Exotic Quantum Phase Group, Institute of Physics, Budapest University of Technology and Economics, H-1521 Budapest, Hungary.}
\affiliation{Department of Physics, University of Oradea, Oradea, 410087, Romania} 
\author{G. Zar\'and}
\affiliation{BME-MTA Exotic Quantum Phase Group, Institute of Physics, Budapest University of Technology and Economics, H-1521 Budapest, Hungary.}
\affiliation{Freie Universit\"at Berlin, Fachbereich Physik, Arnimallee 14, D-14195 Berlin, Germany.} 
\author{P. Simon} 
\affiliation{Laboratoire de Physique des Solides, Univ. Paris-Sud, CNRS, UMR 8502, F-91405 Orsay Cedex, France.}  
\author{H. Bouchiat} 
\affiliation{Laboratoire de Physique des Solides, Univ. Paris-Sud, CNRS, UMR 8502, F-91405 Orsay Cedex, France.}  
\author{R. Deblock} 
\affiliation{Laboratoire de Physique des Solides, Univ. Paris-Sud, CNRS, UMR 8502, F-91405 Orsay Cedex, France.}  

\begin{abstract}
The current emission noise of a carbon nanotube quantum dot in the Kondo regime is measured at frequencies $\nu$ of the order or higher than the frequency associated with the Kondo effect $k_B T_K/h$, with $T_K$ the Kondo temperature. The carbon nanotube is coupled via an on-chip resonant circuit to a quantum noise detector, a superconductor-insulator-superconductor junction. We find for $h \nu \approx k_B T_K$ a Kondo effect related singularity at a voltage bias $eV \approx h \nu $, and a strong reduction of this singularity for $h \nu \approx 3 k_B T_K$, in good agreement with theory. Our experiment constitutes a new original tool for the investigation of the non-equilibrium dynamics of many-body phenomena in nanoscale devices. 
\end{abstract}

\maketitle  

How does a correlated quantum system react when probed at frequencies comparable to its intrinsic energy scales ? Thanks to progress in on-chip detection of high frequency electronic properties, exploring the non-equilibrium fast dynamics of correlated nanosystems is now accessible though delicate. In this respect, the Kondo effect in quantum dots is a model many-body system, where the spin of the dot is screened by the contacts' conduction electrons below the Kondo temperature $T_K$ \cite{goldhaber98,cronenwett98,nigard00}. The Kondo effect can then be probed at a single spin level and in out-of-equilibrium situations. It leads to a strong increase of the conductance of the quantum dot at zero bias due to the opening of a spin degenerate conducting channel, the transmission of which can reach unity. This effect has been extensively studied by transport and noise experiments in the low frequency limit \cite{meir02,sela06,golub06,gogolin06,mora08,delattre09,zarchin08,yamauchi11}. However the noise in the high frequency limit has not been explored experimentally despite the fact that it allows to probe the system at frequencies of the order or smaller than $k_B T_K/h$ characteristic of the Kondo effect \cite{nordlander99}. In this letter we present the first high frequency noise measurements of a carbon nanotube quantum dot in the Kondo regime. We find for $h \nu \approx k_B T_K$ a Kondo effect related singularity at a voltage bias $eV \approx h \nu $, and a strong reduction of it for $h \nu \approx 3 k_B T_K$. These results are compared to recent theoretical predictions.
 
\begin{figure} 
  \begin{center} 
		\includegraphics[width=8.5cm]{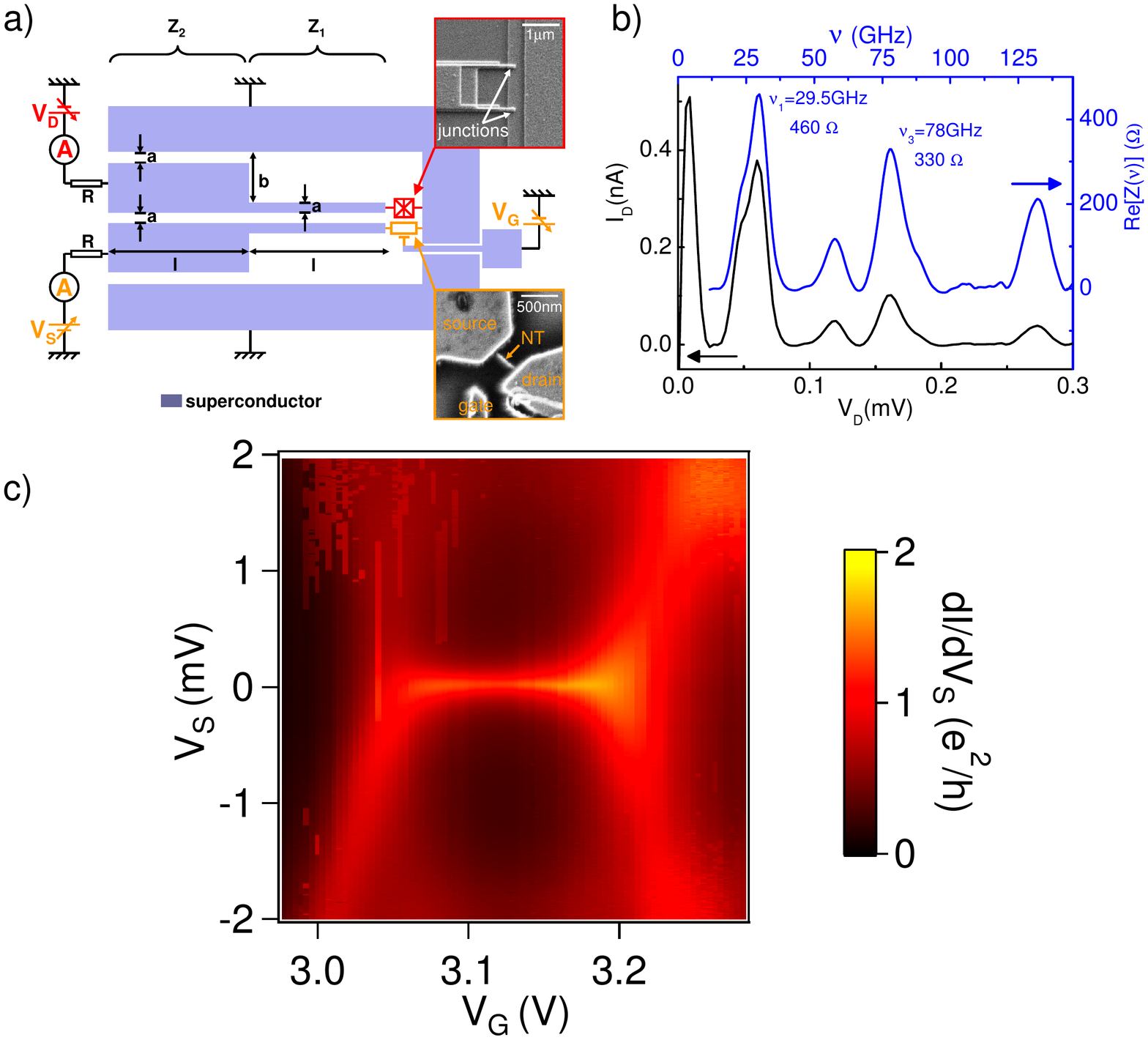} 
	\end{center} 
  \caption{(a) Sketch of the sample: a carbon nanotube (bottom
    electron microscope picture) is coupled to a
    superconductor-insulator-superconductor (SIS) junction (top
    electron microscope picture), used as a quantum detector, by a
    superconducting resonant circuit. This circuit is constituted by
    two transmission lines, placed close to one another, made of
    aluminum and terminated by on-chip Pd resistors. The SIS junction
    is fabricated by shadow angle evaporation. The carbon nanotube is
    CVD grown, connected with Palladium contacts and side-gated. (b)
    Lower curve : $I(V)$ of the detector in the subgap region. Upper
    curve: The real part of the impedance seen by the detector
    exhibits several resonances. (c) Differential conductance
    $dI/dV_S$ of the carbon nanotube as a function of voltage bias
    $V_S$ and gate voltage $V_G$. It exhibits a Kondo ridge for gate
    voltage between 3.05 and 3.20 V with an increase of conductance at
    zero bias.}  
  \label{fig1} 
\end{figure}

The high frequency current fluctuations are measured by coupling the carbon nanotube (CNT) to a quantum noise detector, a Superconductor-Insulator-Superconductor (SIS) junction, \textit{via} a superconducting resonant circuit (see figure~\ref{fig1}a). This allows us to probe  the emission noise of the CNT at the resonance frequencies of the coupling circuit (29.5 GHz and 78 GHz) by measuring the photo-assisted tunneling current through the detector \cite{basset10}. The probed sample consists of two coupled coplanar transmission lines. One line is connected to the ground plane via a carbon nanotube and the other via a superconducting tunnel junction of size $240\times150$ nm$^2$ (figure \ref{fig1}). Each transmission line consists of two sections of same length $l$ but different widths, thus different characteristic impedances $Z_1 \approx 110 \Omega$ and $Z_2 \approx 25 \Omega$ (figure \ref{fig1}a). Due to the impedance mismatch, the transmission line acts as a quarter wavelength resonator, with resonances at frequencies $\nu_n=nv/4l=n\nu_1$, with $v$ the propagation velocity and $n$ an odd integer \cite{basset10}. The two transmission lines are close to one another to provide a good coupling at resonance and are terminated by on-chip Pd resistors. The junction has a SQUID geometry to tune its critical current with a magnetic flux. The carbon nanotube (CNT) is first grown by chemical vapor deposition on an oxidized undoped silicon wafer \cite{kasumov07}. An individual CNT is located relative to predefined markers and contacted to palladium leads using electron-beam lithography. The junction and the resonator are then fabricated in aluminum (superconducting gap $\Delta =182 \mu$eV). A nearby side-gate allows to change the electrostatic state of the nanotube. The system is thermally anchored to the cold finger of a dilution refrigerator of base temperature 20 mK and measured through low-pass filtered lines with a standard low frequency lock-in amplifier technique.

To characterize the CNT-quantum dot, we  first measure its differential
conductance $dI/dV_S$ as a function of dc bias voltage $V_S$ and gate
voltage $V_G$ (figure~\ref{fig1}c). For a gate voltage between 3.05 and
3.2V the CNT's conductance at zero bias strongly increases, a signature of the Kondo effect. The half width at half maximum (HWHM) of the Kondo ridge yields the Kondo temperature $T_K = 1.4$K in
the center of the ridge \cite{goldhaber98}. This value is also consistent
with the temperature dependence of the zero bias conductance. The
Kondo temperature is related to the charging energy $U$ of the CNT
quantum dot, the coupling $\Gamma$ to the electrodes and the position
$\epsilon$ of the energy level measured from the center of the Kondo
ridge, according to Bethe-Ansatz~\cite{tsvelick83,bickers87}:  
\begin{equation} 
	T_K=\sqrt{U\Gamma/2}\exp\left[- \frac{\pi}{8U\Gamma}|4\epsilon^2-U ^2|\right] \;. 
	\label{TK} 
\end{equation} 
From $U=2.5$meV, deduced from the size of the Coulomb diamond, and $T_K=1.4$K, we obtain $\Gamma=0.51$meV. The asymmetry $A=(\Gamma_L-\Gamma_R)/(\Gamma_L+\Gamma_R)=0.67$ of the contacts is deduced from the zero bias conductance.  
 
To characterize the superconducting resonant circuit which couples the
detector junction to the CNT, we measure the subgap $I(V)$
characteristic of the junction which depends on the impedance of its
electromagnetic environment \cite{ingold92}. In the case of a
superconducting transmission line resonator \cite{basset10,holst94},
resonances appear in the subgap region $V_D<2\Delta/e$ due to the
excitation of the resonator modes by the ac Josephson effect
\cite{barone82}. These resonances are related to the real part of the
impedance $Z(\nu)$ seen by the junction:  
\begin{equation} 
	I(V_D)=Re[Z(2eV_D/h)] \, I_C^2/2V_D\;,  
	\label{Ijj} 
\end{equation} 
with $I_C=\pi \Delta/(2eR_N)$ the critical current \cite{barone82}, $R_N=28.6 k\Omega$ the normal state resistance of the junction and $\Delta$ the superconducting gap of the electrodes. Equation~\ref{Ijj} accounts for the effect of the
electromagnetic environment on the tunneling of Cooper pairs through the Josephson junction \cite{ingold92}. Figure \ref{fig1}b shows the $I(V)$ of the junction in the subgap region for $I_C$
maximized with magnetic flux. The subgap resonances thus yield via equation \ref{Ijj} $Re[Z(\nu)]$ (Fig. \ref{fig1}b) which is peaked at frequencies $\nu_1=29.5$ and $\nu_3=78$GHz. Using the height and width of the resonance peaks of $Re[Z(\nu)]$, we infer the coupling between the junction and the CNT \cite{basset10}. We then translate a photo-assisted tunneling (PAT) quasi-particles current measurement into a current emission noise measurement for the frequencies $\nu_1$ and $\nu_3$. The ratio $r_n$ between the measured PAT current through the detector and the current emission noise of the CNT at a given resonance frequency is estimated as follows. $r_n$ is given at the resonant frequency $\nu_n$ by $e^2 |Z_t(\nu_n)|^2 \delta\nu_n/(h\nu_n)^2 I_{QP,0}(V_D+\frac{h\nu_n}{e})$ with $Z_t$ the transimpedance of the coupling circuit, defined as the ratio between the voltage fluctuations across the detector and the current fluctuations through the source, $\delta\nu_n$ the width of the resonance peak and $I_{QP,0}(V_D)$ the \textit{I-V} characteristic of the detector \cite{basset10}. This value has been calibrated in a
previous experiment with the same design as the one used in the present work \cite{basset10}. $r_n$ is then calculated using the ratio of the calibrated sample corrected according to the square area under the corresponding peak of $Re[Z(\nu)]$ (figure \ref{fig1}b), the value of $\nu_n$, the superconducting gap and the tunnel resistance of the detector junction.  
\begin{figure} 
  \begin{center} 
		\includegraphics[width=8.5cm]{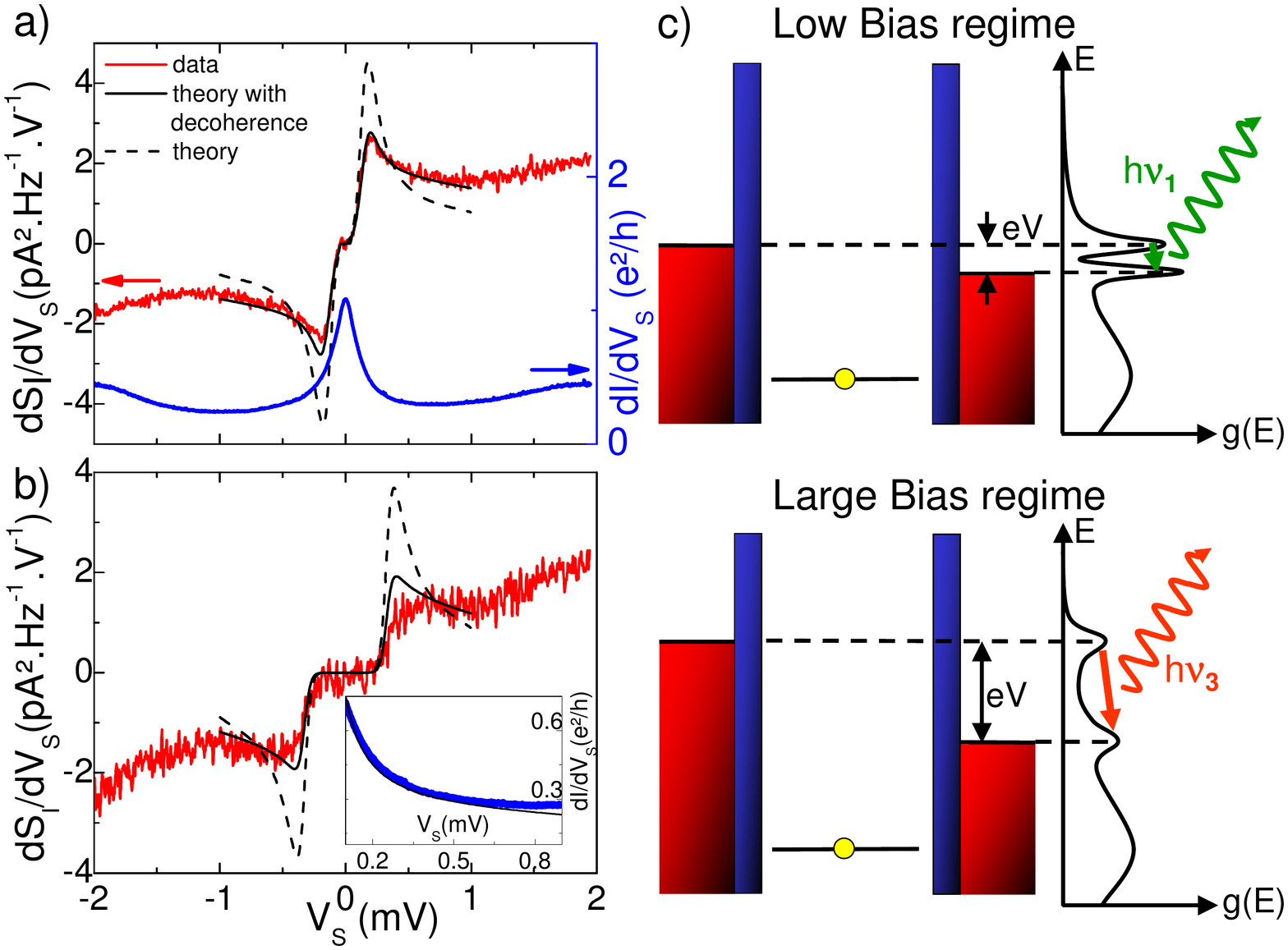} 
	\end{center} 
  \caption{(a) Derivative of the current noise and differential
    conductance of the CNT versus $V_S$ in the center of the Kondo
    ridge. Left axis : measured $dS_I/dV_S$ (in red) at $29.5$GHz as a
    function of bias voltage $V_S$. The black solid line corresponds to
    the calculated $dS_I/dV_S$ with a fitted decoherence rate 
    incorporating both intrinsic and extrinsic decoherence (see
    text), while  the dashed line is the
    theoretical prediction with the intrinsic
    decoherence rate. Right axis : Differential conductance $dI/dV_S$
    of the CNT in units of $e^2/h$. (b) Same data
    at $78$GHz. Inset : measured conductance,
    symmetrized with respect to the bias voltage, in units of
    $e^2/h$. The black solid line correspond the the FRG calculation
    of the conductance with the \textit{same} fitted decoherence
    rate as the one used for the noise measurements. (c) Schematic picture of a quantum dot in the
    out-of-equilibrium Kondo regime together with the density of
    states for two distinct bias voltages. When the quantum dot is
    voltage biased, the Kondo resonance splits in two with a splitting
    given by the applied bias voltage. This leads to an increase of
    the emission at frequency $eV_S=h\nu$. The amplitude of the
    resonance peaks, and thus the emission noise at $eV_S=h\nu$, can
    be reduced by decoherence effects induced by the applied bias
    voltage.}  
  \label{fig2} 
\end{figure}

To measure the quantum noise of the CNT, we modulate its bias voltage $V_S$ and monitor the modulated part of the PAT current through the detector for a given detector bias voltage $V_D$. $V_D$ selects the frequency range of the measurement \cite{basset10}. We have thus access to the derivative of the PAT current versus CNT bias voltage
$dI_{PAT}/dV_S$ at a given frequency. Using the previously estimated coupling coefficient, we translate this quantity into the derivative of the current noise $S_I$ at one of the resonance frequencies versus
$V_S$, $dS_I/dV_S$. This quantity is plotted in the center of the Kondo ridge, \textit{i.e.} $\epsilon=0$ at two frequencies (Figure \ref{fig2}a and b). For each frequency, the data exhibit a region close to $V_S=0$ where $dS_I/dV_S=0$. This corresponds to $|e V_S| < h \nu$, where the system does not have enough energy to emit noise at a frequency $\nu$. The observation of this zero noise region is a strong evidence that we are indeed only measuring the emission noise of the CNT. For $|e V_S| > h \nu$ the system emits noise at $\nu$. For the first resonance  frequency $\nu_1=29.5$GHz, with $h \nu_1 \approx k_B T_K$, the measured derivative of the noise shows a singularity for bias voltages close to the measured frequency. At higher bias voltages $dS_I/dV_S$ is much smoother. For $h \nu_3 \approx 2.7 k_B T_K$ the previous singularity is nearly absent and $dS_I/dV_S$ versus $V_S$ is practically flat.  
 
The high frequency noise of quantum dots in the Kondo regime has been
studied theoretically at equilibrium using the numerical
renormalization group (NRG) technique \cite{sindel05}.  Non-equilibrium
results for the finite-frequency 
noise are theoretically much more demanding. They were obtained only 
for peculiar values of parameters (strongly anisotropic exchange
couplings) of the Kondo problem using bosonization methods
\cite{schiller98}, and by using  non-equilibrium real time
renormalization group approaches \cite{korb07,moca11}. The latter
approaches  assume  $h \nu, eV_S \gg k_B T_K^{\rm RG}$, with $T_K^{\rm
  RG}$ the  Kondo temperature defined from the renormalization
group. Importantly, $T_K^{\rm RG}$ differs from $T_K$ (defined
experimentally as the HWHM of the differential
conductance) by a numerical factor,  which has to be
determined (see below). Here we employ the real time functional
renormalization group (FRG) approach developed in Ref.~\cite{moca11}
to compute the non-equilibrium frequency-dependent noise  and compare
it to the experimental results. We perform the non-equilibrium
calculations using  the Kondo Hamiltonian, given by~:  
\begin{equation}
H_K=\frac{1}{2}\sum_{\alpha,\beta=L,R} \sum_{\sigma,\sigma'}
j_{\alpha\beta}\;
\psi_{\alpha\sigma}^\dagger \;\mathbf{S}\cdot {\boldsymbol \sigma}_{\sigma\sigma'}\;
\psi_{\beta\sigma'}\;.
\label{eq:H_K}
\end{equation}
Here the $j_{\alpha\beta}$ denote the Kondo couplings, $\alpha,\beta$
are indices for the left (L) and right (R) leads, $\boldsymbol \sigma$
stands for the three Pauli matrices, and the operator
$\psi_{\alpha\sigma}$ destroys an electron of spin $\sigma$  in lead
$\alpha\in\{L,R\}$. We parametrize the dimensionless exchange
couplings $j_{\alpha\beta}$ as $j_{\alpha\beta}= j \; v_\alpha
v_\beta$,  with the  factors $\{v_L,v_R\} =
\{\cos(\phi/2),\sin(\phi/2)\}$ accounting for the asymmetry of the
quantum dot, $\cos(\phi) = A$, and $\phi$
related to the $T=0$ conductance as $G (T=0) = (2e^2/h)\; \sin^2(\phi)$. 

The Kondo Hamiltonian assumes that charge fluctuations in the CNT
quantum dot are frozen. Therefore, the theoretical results based on
(\ref{eq:H_K}) can and shall be compared with experimental ones only
for bias voltages $V_S\ll U/e$.  As a first step, to determine the
ratio $T_K^{\rm RG}/T_K$, we  computed the equilibrium conductance
by using NRG \cite{NRG} and
compared it to experimental data in the center of the Kondo ridge ($\epsilon = 0$). This enabled us to establish 
$T_K\approx 3.7 \;T_K^{\rm RG}$ (see the appendix). Therefore, the condition  $h\nu \gg k_B T_K^{\rm RG}$
for our FRG approach to apply is certainly met for the frequency
$\nu_3$,  and still reasonably satisfied for $\nu_1$. 

Within the Kondo model, we can express the Fourier transform of the emission noise $S_I$ as :
\begin{equation}
S_I(V_S,\nu)=
\frac{e^2}{h}\; k_B T_K^{\rm RG}\;
s\left(\frac{eV_S}{k_BT_K^{\rm RG}},\frac{h\nu}{k_BT_K^{\rm RG}},\frac{T}{T_K^{\rm RG}},A\right),
\label{eq:SV}
\end{equation}
where $s$ is a dimensionless function, which we calculate by solving numerically the FRG
equation (see appendix). Since the measurement temperature satisfies
$T\ll T_K$, we have taken $T=0$ in the calculations and checked that 
a finite but small temperature does not affect our
results. Note that no fitting parameter has been included at this
level, since the asymmetry parameter $A$ and $T_K^{\rm RG}\approx
0.38 \;{\rm K}$ were extracted from the experimental
data. The dashed lines in figure~\ref{fig2}a and b show
the calculated  $dS_I/dV_S$ curves for frequencies $\nu_1=29.5$GHz
and $\nu_3=78$GHz, respectively. The computed curves are only shown in
the  bias range $|V_S|<1$ mV, where the Kondo Hamiltonian in
equation (\ref{eq:H_K}) is appropriate to  describe  the physics of the CNT
quantum dot.  For both frequencies, the theoretical curves exhibit
sharp singularities at $eV_S=h\nu$, much more pronounced than the 
experimental ones. This especially holds for the resonance frequency,
$\nu_3=78$GHz, where the resonance is almost completely absent
experimentally.   The singularity at the threshold  $eV_S\approx h\nu$
is related to the existence of two Kondo resonances associated with
the Fermi levels of the two contacts. Inelastic transitions between
them lead to an increase of $dS_I/dV_S$ for frequencies
corresponding to the energy separation, 
 $h\nu \approx e V_S$ (figure~\ref{fig2}c).  
\begin{figure} 
  \begin{center} 
		\includegraphics[width=8.5cm]{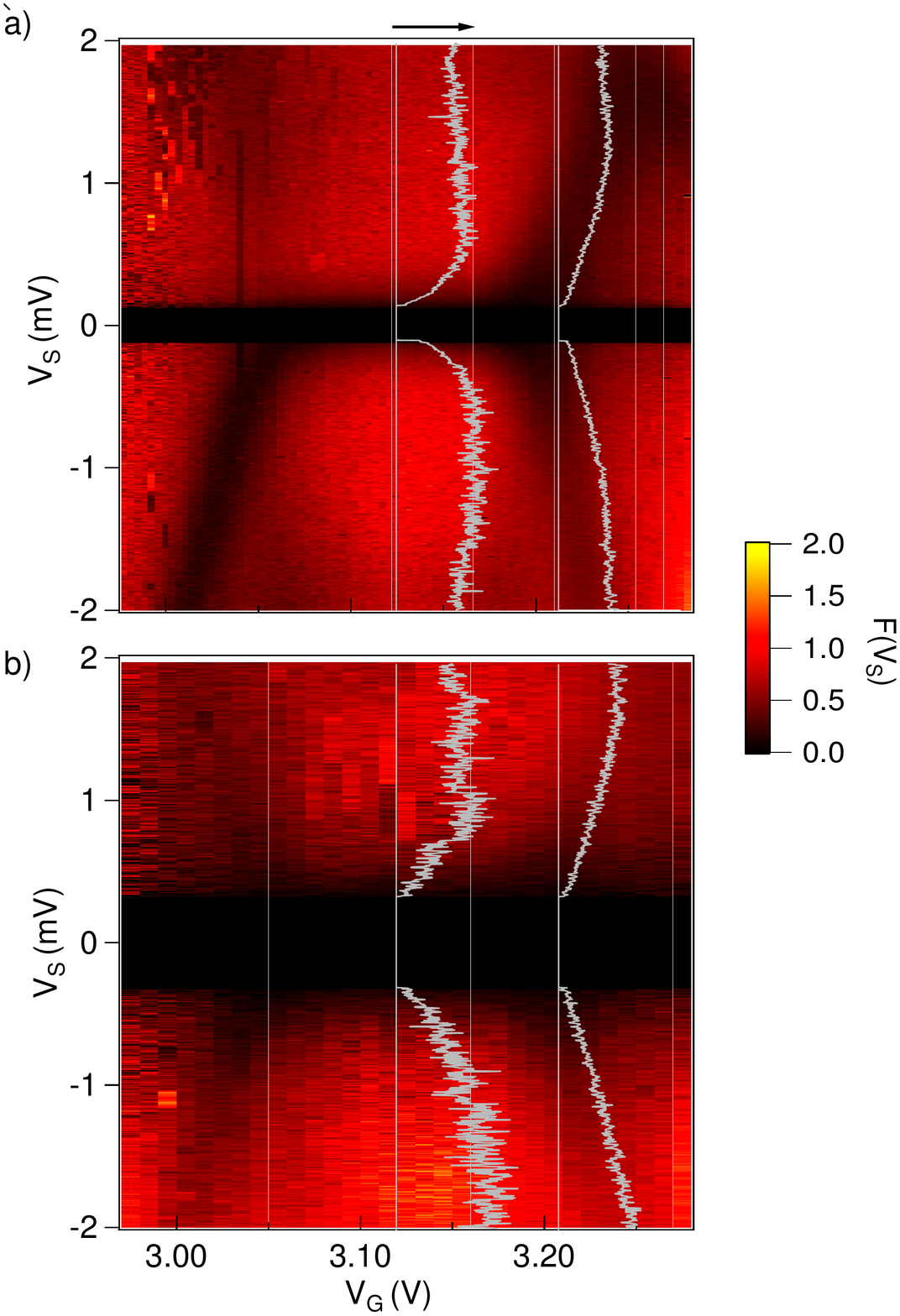} 
	\end{center} 
  \caption{$F(V_S)=[dS_I/dV_S]/[e \, dI/dV_S(V_S-h\nu/e)]$ as a function
    of the bias voltage $V_S$ and the gate voltage $V_G$ at
    $\nu_1=29.5$GHz (a) and $\nu_3=78$GHz (b). $F(V_S)$ is arbitrarily
    fixed to zero for $e|V_S|<h\nu$. The gray curves on top of the
    colorplot correspond to the bias voltage dependence of $F(V_S)$ at
    $V_G=3.12$V and $V_G=3.21$V. The black arrow indicates a value of
    the $F(V_S)$ equal to 1. When the conductance is low, $F(V_S)$ is
    close to one while it is reduced in the highly conducting regions.}  
  \label{fig3} 
\end{figure}  

To compute the dashed curves in figure \ref{fig2}, an intrinsic spin
decoherence time $\tau_S$ induced by the large bias was included and
calculated self-consistently in the FRG approach \cite{paaske04} (see appendix).  The decoherence of the Kondo effect induced by a large
dc voltage bias is a well-known feature  which has indeed 
been measured  \cite{francesci02,leturcq05}, and has been predicted to lead to a
strong reduction of the Kondo resonance due to inelastic
processes~\cite{paaske04,monreal05,roermund10}. Since the singularity
in the noise at $h\nu \approx e V_S$ is associated with the transitions between the two Kondo
resonances pinned at the Fermi levels of the contacts, this
singularity is also affected by decoherence. 

However, as shown in  figure~\ref{fig2}, the computed intrinsic
decoherence time is insufficient to explain the experimentally
observed suppression of the peak in $dS_I/dV_S$.  Therefore, we
incorporated a voltage-dependent spin relaxation rate
in our calculations, $\tau_S^{-1}(V_S)$, which includes 
external decoherence. The consistency of this approach can be checked
against the experiments: a {\em single} choice of $\tau_S^{-1}(V_S)$ must
simultaneously reproduce the voltage dependence of the differential
conductance through the dot $dI/dV_S(V_S)$, and those of the
$\nu_1=29.5$ GHz and $\nu_1=78$ GHz  noise spectra,
$dS_I/dV_S(V_S)$. Furthermore, $\tau_S^{-1}$ should be suppressed for
$V_S < T_K^{\rm RG}$. We found that a bias-dependent decoherence rate
of the form $h/\tau_S\approx \alpha\; k_B T^{\rm RG}_K {\rm
  atan}(\beta e V_S/k_B T^{\rm RG}_K)$ (similar in shape to the
calculated intrinsic spin relaxation rates),  with $\alpha=14$ and $\beta =
0.15$ satisfied all criteria above. The continuous lines in
figure~\ref{fig2} show  the $dS_I/dV_S$ curves computed with
this form of $h/\tau_S$, and fit fairly well the experimental data for
both resonator frequencies. As a final consistency check, we also
computed the differential conductance through the dot (taking into
account the above form of $\tau_S^{-1}$) and compared it to the
measured $dI/dV_S$ curves. A very good agreement is found  without any
other adjustable parameter in the voltage-range 
$V>0.1mV$,  where the FRG approach is appropriate (Inset of figure \ref{fig2}b).

From the theoretical fits we infer that the experimentally observed
noise spectra and differential conductance  can be understood in terms
of a decoherence rate, which  is about a factor of $2$ larger
than the  theoretically computed intrinsic rate (see appendix). One possibility for
this discrepancy is that the experimentally observed decoherence is 
intrinsic, and  FRG - which is a perturbative approach - underestimates the spin relaxation rate in this 
regime (which is indeed almost out of the range of perturbation
theory). Another possibility is that the  
experimental set-up leads to additional decoherence. 

The experiment also allows to draw a complete map of the noise in the
region of the Kondo ridge. We define $F(V_S)=[dS_I/dV_S(V_S)]/[e \, dI/dV_S(V_S-h\nu/e)]$, \textit{i.e.} the ratio
of the derivative of the noise to the differential conductance
shifted in voltage by an amount corresponding to the measured
frequency. For both linear and non linear systems with energy
independent transmission at low temperature this quantity is equal to
the Fano factor \cite{blanter00}. We have plotted $F(V_S)$ for $\nu_1=29.5$GHz
(figure \ref{fig3}a) and $\nu_3=78$GHz (figure \ref{fig3}b). For $|eV_S| <
h \nu$, where the emission noise is zero, $F(V_S)$ is arbitrarily fixed
to zero. For both frequencies the noise is found to be sub-poissonian,
with $F(V_S)$ close to one in the poorly conducting regions and a strong
decrease of $F(V_S)$ along the conducting regions. This is qualitatively consistent with the reduction of the Fano factor for a conducting
channels of transmission close to one. This result has to be contrasted with back scattering noise measurements in the Kondo regime at low frequency and low bias voltage where the Fano factor was found to be higher than one \cite{zarchin08,yamauchi11}.

In conclusion we have measured the high frequency current fluctuations
of a carbon nanotube quantum dot in the Kondo regime by coupling it to
a quantum detector \textit{via} a superconducting resonant circuit. We
find that the noise  exhibits strong resonances when the voltage
bias is of the order of the measurement frequency in  good agreement with  theory
provided that an additional  decoherence rate is included which prevents the full
formation of the out of equilibrium Kondo resonances. Our experiment constitutes a new original tool for the investigation of the non-equilibrium dynamics of many-body phenomena in nanodevices.

We thank M. Aprili, S. Gu\'eron, M. Ferrier, M. Monteverde and J. Gabelli for
fruitful discussions. This work has benefited from financial support of ANR under Contract DOCFLUC (ANR-09-BLAN-0199-01) and C'Nano Ile de France (project HYNANO), the EU-NKTH GEOMDISS project, OTKA research Grants No. K73361 and No. CNK80991 and the French-Romanian grant DYMESYS (ANR 2011-IS04-001-01 and contract PN-II-ID-JRP-2011-1).

\section*{Appendix : Relation between $T_K$ and $T_K^{RG}$}

To perform the theoretical calculations, one first needs to find the Kondo 
temperature. However, the Kondo temperature is defined only 
up to a prefactor, and its value also 
depends slightly  on the physical quantity from which it is defined. 
Our experimental Kondo temperature, $T_K$, is defined
as $T_K = e\;\Delta V_S/k_B$, with $\Delta V_S$
 the half-width at half maximum of the measured $G(V_S)=dI/dV_S$ 
curves.
In the FRG calculations, on the other hand,  it is defined as 
a scale (frequency),  $k_B T_K^{\rm RG}\equiv \hbar \omega_K^{\rm RG}$, 
where the so-called leading 
logarithmic calculations yield a divergent interaction vertex at 
$T=0$ temperature. It can, however, also be defined as 
the temperature, $T_K^{\rm T-dep}$, at which the linear conductance drops 
to half of its $T=0$ temperature value. The ratios of all these 
Kondo temperatures are just universal numbers (apart from a 
possible but presumably small dependence of $\Delta V_S$ on the anisotropy, $A$).

To determine the ratio of $T_K^{\rm T-dep}$ and $k_B T_K^{\rm RG}$, 
we performed  numerical renormalization group calculations~\cite{BudapestCode}: 
We computed the full $G(T)$ curve, extracted from it the 
width $T_K^{\rm T-dep}$, and for the same parameters, we also computed 
$T_K^{\rm RG}=\hbar \omega_K^{\rm RG}/k_B $ 
from the high-frequency tail of the so-called composite 
fermions' spectral function, scaling as $\approx C/\ln^2(\omega/\omega_K^{\rm RG})$
at large frequencies. In this way, we obtained a ratio
$$
T_K^{\rm RG}/T_K^{\rm T-dep} \approx \; 0.3.
$$
The ratio $T^{\rm T\;dep}_K/T_K$ was then determined from experimental 
data \cite{Kouwenhoven}, giving 
$$
T_K^{\rm T-dep}/T_K \approx 1.2\;.
$$
The previous equations yield the ratio, $T_K^{\rm RG}/T_K\approx 0.36$, 
used in our calculations and quoted in the main text. 

\section*{Appendix : Summary of the functional renormalization group approach}

In this work we used the functional renormalization 
group approach developed in Ref.~\cite{moca11}, an 
extension of the formalism of Ref.~\cite{rosch}. 
In this approach, formulated at the level of non-equilibrium action, 
a short time cut-off $a$ is introduced, and increased in course of the 
renormalization group (RG) procedure to eliminate the high-energy 
degrees of freedom. This procedure yields a retarded interaction 
($j_{\alpha\beta}\to g_{\alpha\beta}(t-t',a)$), 
whose cut-off dependence is  described by the  differential equation,
\begin{equation}
\frac {\rm d  {\bf g} (\omega, a)}{{\rm d}\ln a}=
{\bf g} (\omega,a)\; {\bf q}(\omega,a)\;{\bf g} (\omega,a)\;.
\label{eq:scaling_g}
\end{equation}
Here we introduced the matrix notation, 
$g_{\alpha\beta}(\omega,a)\to {\bf g}(\omega,a)$ for the Fourier 
transform of the 
retarded interaction, and the matrix $ {\bf q} (\omega,a)$
denotes a cut-off function. In our calculations
we have not approximated  this latter 
 with  $\Theta$-functions, as in Ref.~\cite{moca11},  
but used a function corresponding to the real time 
propagators of Ref.~\cite{moca11}, also incorporating 
the effect of an exponential decay rate $\Gamma =1/\tau_S$. 
\begin{figure}
  \begin{center}
		\includegraphics[width=8.5cm]{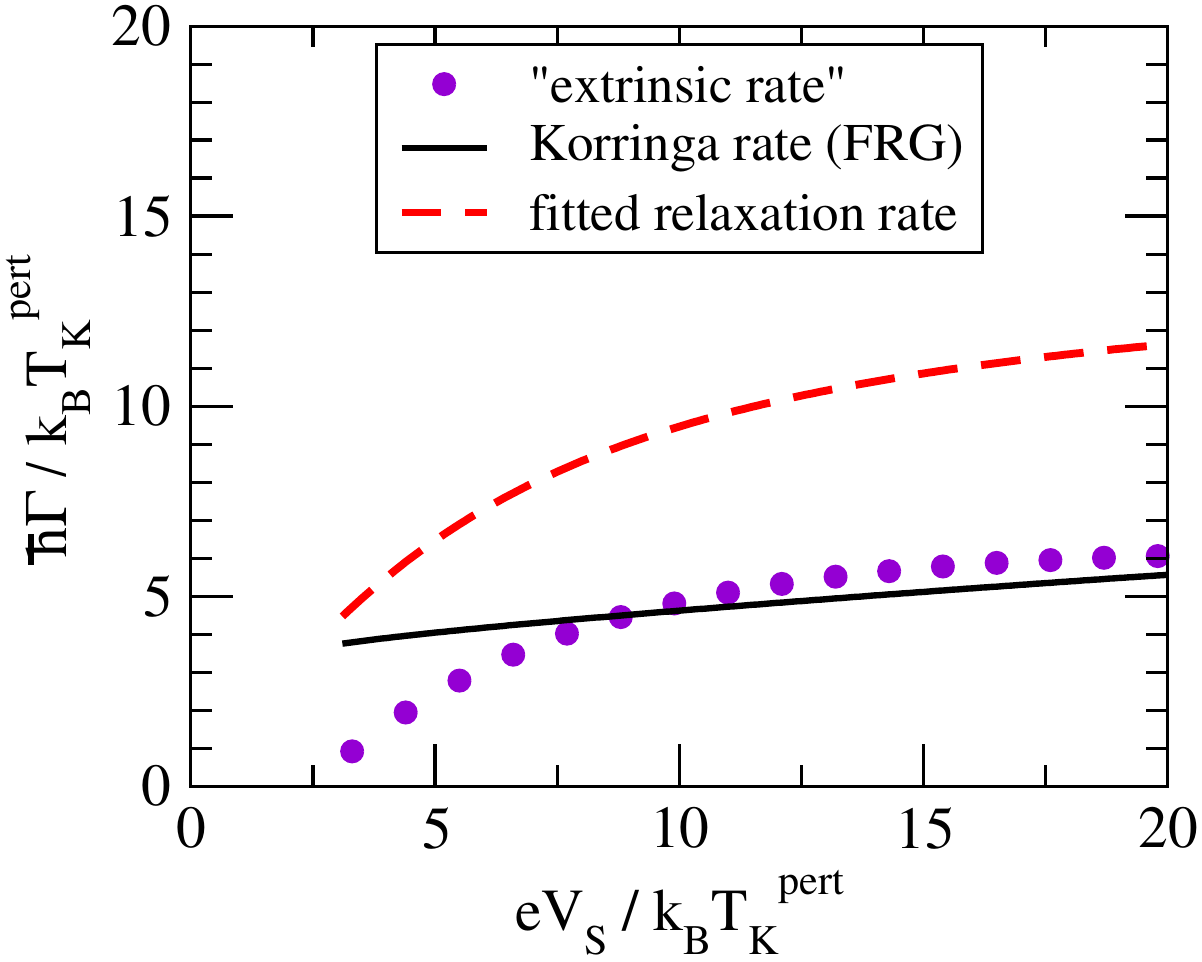}
  \end{center}
  \caption{Voltage dependence of the fitted and intrinsic relaxation rates, and 
their difference, termed as extrinsic spin relaxation rate
in the voltage range, $V_S\ge 0.1\; {\rm mV}$. 
The intrinsic rate was calculated by FRG, while the form and 
overall magnitude  of the extrinsic rate were inferred from 
the differential conductance data.}
\label{rates}
\end{figure}

The voltage-dependent decay rate $\Gamma(V_S)$ has an intrinsic 
part, $\Gamma_{\rm intr}(V_S)$, as well as an
external contribution, $\Gamma_{\rm ext}(V_S)$. The former contribution
can be identified as the Korringa spin relaxation rate, and can be
expressed as 
\begin{eqnarray}
\Gamma_{\rm intr} &=& \pi \sum_{\alpha, \beta=L,R} \int {\rm d}\omega\,
 g_{\alpha\beta}(\omega)\;g_{\beta \alpha}(\omega)\; \nonumber \\
 && f(\omega-\mu_\alpha)\;(1-f(\omega-\mu_\beta))\;,
\end{eqnarray}
with $g_{\alpha\beta}(\omega)$ the vertex functions in the limit
$a\to \infty$, $\mu_\alpha$ the electro-chemical potentials of the leads, 
and $f$ the Fermi function. 
The 
rate  $\Gamma_{\rm intr}(V_S)$, as computed by FRG is shown in Fig.~\ref{rates}. 
Rather surprisingly, in the cross-over regime, 
$eV_S\sim k_B T_K^{\rm pert}$, 
it almost saturates, and only weakly depends on the bias.
For a comparison, figure~\ref{rates} also shows the 
total fitted relaxation rate, $\Gamma(V_S)=1/\tau_S(V_S)$  needed to reproduce 
the differential conductance  and noise  data. 
It is not very far from the calculated intrinsic 
contribution, but it is above the latter, and it apparently 
includes some extrinsic spin relaxation, too.

In the formalism of Ref.~\cite{moca11}, the current operator and the 
current vertex  are also renormalized during the RG procedure, and 
also become non-local. However, the current vertex, $\mathbf{V}$, has a more complicated 
structure than the interaction vertex, and possesses two non-trivial 
time arguments, $g_{\alpha\beta}(\omega)\leftrightarrow {  V_{\alpha\beta}}(\omega_1,\omega_2)$.
The evolution of $ \mathbf{  V}$ under the RG is described by a differential 
equation similar to equation~\eqref{eq:scaling_g}  
(see reference~\cite{moca11} for the details). The noise spectrum, i.e., 
the Fourier transform of the current-current correlation function 
can then be expressed as a double integral of this
retarded  current vertex taken in the limit $a\to \infty$.
The  emission  noise, $S_e(\omega)$,  e.g., can be expressed as 
 \begin{eqnarray}
S_e(\omega) = \frac{e^2} 2\;
S(S+1)&&\int \frac{{\rm d} \tilde \omega}{2\pi}\;{\rm Tr}\{
{\bf V}(\tilde\omega_+,\tilde\omega_-)
 {\bf G}^{>}(\tilde\omega_-) \nonumber \\
 &&{\bf V}(\tilde\omega_-,\tilde\omega_+)  {\bf G}^{<}(\tilde\omega_+)
\}\;,
\end{eqnarray}
with $\tilde\omega_\pm = \tilde \omega \pm \frac \omega 2$ and $S=1/2$.
Here the trace refers to the labels $\alpha,\beta = L/R$,
and the bigger and lesser Green's functions are
given as $ G_{\alpha\beta}^{>/<}(\omega) = \pm i\;2\pi\;
\delta_{\alpha\beta}\;f(\pm(\omega-\mu_\alpha))$.

\end{document}